\documentclass[twocolumn,prb]{revtex4}
\usepackage{graphicx}
\usepackage{epsfig}
\usepackage{dcolumn}
\usepackage{bm}
\usepackage{amsmath}

\begin{document}
\title{Radiation-induced current in quantum wires with side-coupled nano-rings}
\author{Yuriy V. Pershin}
\email{pershin@pa.msu.edu}
\author{Carlo Piermarocchi}
\affiliation{Department of Physics and Astronomy, Michigan State
University, East Lansing, Michigan 48824-2320, USA}

\begin{abstract}
Photocurrent generation is studied in a system composed of a quantum wire with
side-coupled quantum rings. The current generation results from the interplay
of the particular geometry of the system and the use of circularly polarized
radiation. We study the energy-momentum conservation for optical transitions
involving electrons moving forwards and backwards in the wire. Due to the lack
of time-reversal symmetry in the radiation, the optical transitions depend on
the direction of motion of the electrons, leading to a current at zero bias
voltage. The photocurrent increases with the number of rings within a wide
range of physical parameters. A weak non-linear dependence of the current in
the number of rings, related to quantum interference effects, is also
predicted. This geometry suggests a scalable method for the generation sizeable
photocurrents based on nanoscale components.

\end{abstract}

 \maketitle

\section{Introduction}
There is currently a significant theoretical and experimental interest in
radiation-induced currents in confined electron systems.~
\cite{perpierBent,magarill,perpierSO,lit8,lit9,lit10,perpierQR,Mani,Zudov} In
particular, many investigations have focused on novel mechanisms for
photo-current generation based on nano-scale components. We have recently
investigated a mechanism for photocurrent generation in a bent quantum wire by
circularly-polarized radiation.~\cite{perpierBent} The curved geometry provides
the optimal condition for exploiting the broken time-reversal symmetry in the
circularly polarized electromagnetic field. The mechanism of photocurrent
generation in curved structures results from the interplay of circularly
polarized radiation and the peculiar geometry, which yields an effective
time-dependent sliding potential for the electrons in the bent
region.~\cite{magarill} In fact, the radiation induced sliding potential
induces an asymmetry in the scattering coefficients for electrons moving
forwards and backwards and a net current appears at zero bias voltage.
Radiation-induced transitions that reverse the sign of the electron velocity
play a key role in the photocurrent generation. We found that for a realistic
set of parameters the calculated photocurrent in bent quantum wires is
relatively weak, of the order of $\sim 10$pA.

In the present paper, we consider photocurrent generation in a quantum wire
with side-coupled quantum rings. This is a scalable structure with a mechanism
of photocurrent generation similar to the one described above. Circularly
polarized radiation creates a sliding potential in all the quantum rings
coupled to the wire. This potential asymmetrically change the scattering
coefficients through the whole structure. The advantage over a single bent
quantum wire is in the higher sensitivity to the radiation, which is related to
the fact that the electron propagation involves trajectories with multiple
passes through the ring regions. The path in which the electromagnetic field
affects the dynamics is more extended, and the changes in the forward and
backward scattering are more significant. Moreover, since there are many rings
coherently coupled to the electromagnetic field, we are interested in exploring
the non-linear features related to the coherent behavior of the electron in the
many-ring structure.

\begin{figure}[b]
\centering
\includegraphics[angle=270,width=8.5cm]{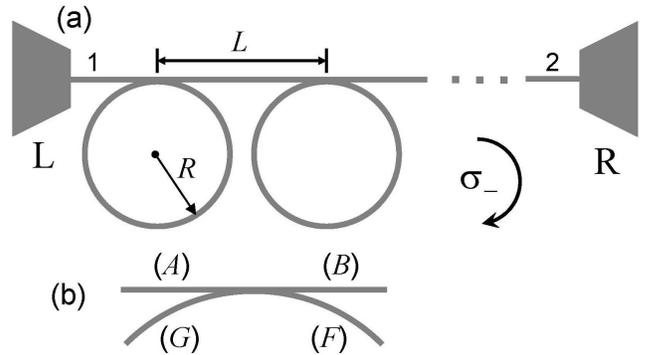}
\caption{(a) Quantum wire with $N$ side coupled quantum rings in the presence
of circularly-polarized radiation. The radiation breaks the symmetry for left-
and right-moving electrons, inducing a photocurrent. (b) Definition of the wave
function amplitudes near a single wire-ring junction.} \label{fig1}
\end{figure}

Fig. \ref{fig1}(a) shows the system under investigation. We consider a chain of
$N$ one-dimensional rings, of radius $R$, in the $x-y$ plane, side-coupled
(from the same side) to a straight quantum wire in the $x$ direction. The
distance between the centers of adjacent rings is $L$. A circularly-polarized
electromagnetic radiation propagates in $z$ direction, perpendicular to $x-y$
plane. The quantum wire is also connected to two electron reservoirs (denoted
by L (left) and R (right)) at zero bias voltage. We are interested in the
photocurrent generated in such a system. We can assume that the quantum wire
and quantum rings are narrow in the transverse directions, so that our
consideration is limited only to electrons within the lowest transverse
sub-band. A radiation acting on the straight segments of the wire will not
generate a current, no matter what polarization is used. We will therefore
neglect the effect of the electromagnetic radiation on the straight segments of
the nanostructure.

We calculate the radiation-induced current in the structure in the ballistic
transport regime. In order to calculate the current, we solve a
radiation-dependent scattering problem for the electrons in the structure using
a transfer-matrix approach. The total transfer-matrix is calculated as a
product of transfer-matrices associated to many smaller segments in the
structure. The transmission properties at the wire-ring junctions are modeled
by a unitary matrix which depend on a parameter describing the strength of
wire-ring coupling. The current is calculated as a function of different
coupling parameters and its scaling with the system size is considered. In a
wide range of photon energies the photocurrent almost linearly increases with
the number of quantum rings attached to the wire, but signatures of non-linear
effects are also observed. The paper is organized as follows. The mechanism of
photocurrent generation is discussed in Sec. \ref{mechn}, and  the calculation
scheme is introduced in the Sec. \ref{sec2}. In Sec. \ref{sec3} we present the
results of calculations, and Sec. \ref{sec4} contains concluding remarks.

\section{Photocurrent generation mechanism \label{mechn}}
We consider the interaction of electrons confined in a circular segment of the
structure with circularly-polarized electromagnetic radiation propagating
perpendicularly to the quantum wire plane. The interaction term in the
Hamiltonian can be written as \cite{perpierBent}
\begin{equation}
-\boldsymbol{dE}=2eRE_0\sin\left( \frac{s}{R}\pm\omega t\right)
\label{intpot},
\end{equation}
where  $s$ is a coordinate along the ring, $\boldsymbol{d}=-e\boldsymbol{r}$ is
the dipole moment operator. The electric field in the radiation is written as
$\boldsymbol{E}=E_0\cos (\omega t)\hat{x}\pm E_0\sin (\omega t)\hat{y}$, where
$E_0$ and $\omega$ are the electric field amplitude and frequency, $\hat{x}$
and $\hat{y}$ are unit vectors in the $x$ and $y$ directions (in the quantum
wire plane), and $\pm$ corresponds to a $\sigma_\pm$ circular polarization.
According to Eq. (\ref{intpot}), the electrons in the circular segment of the
quantum wire (QW) are subjected to a potential that moves forwards or backwards
depending on the helicity of the circularly polarized light.  A similar sliding
potential describes the interaction of electrons with a traveling acoustic
wave.~\cite{galperin} In our case, the effective wavelength of the traveling
wave is determined by the quantum wire curvature and is $2 \pi R$.

\begin{figure}[t]
\centering
\includegraphics[width=6.5cm]{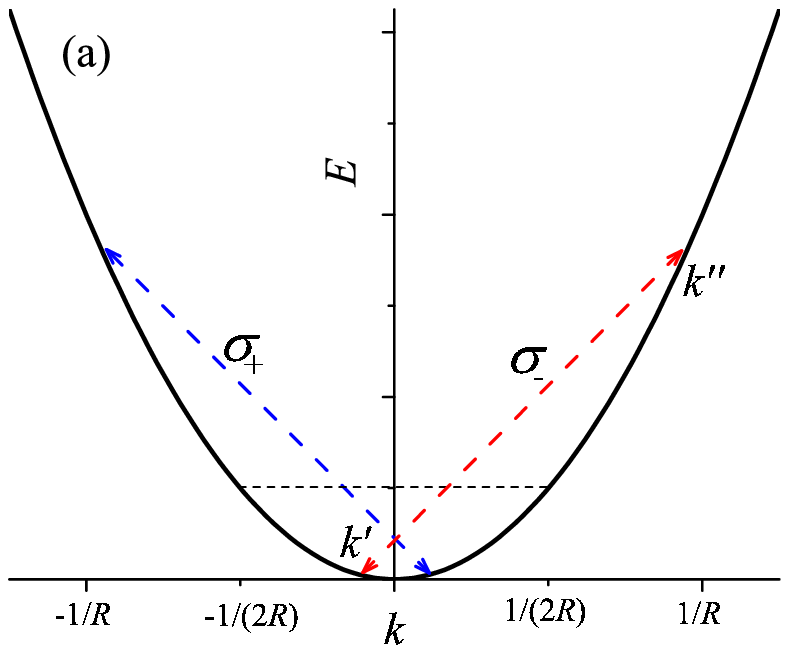}
\includegraphics[width=6.5cm]{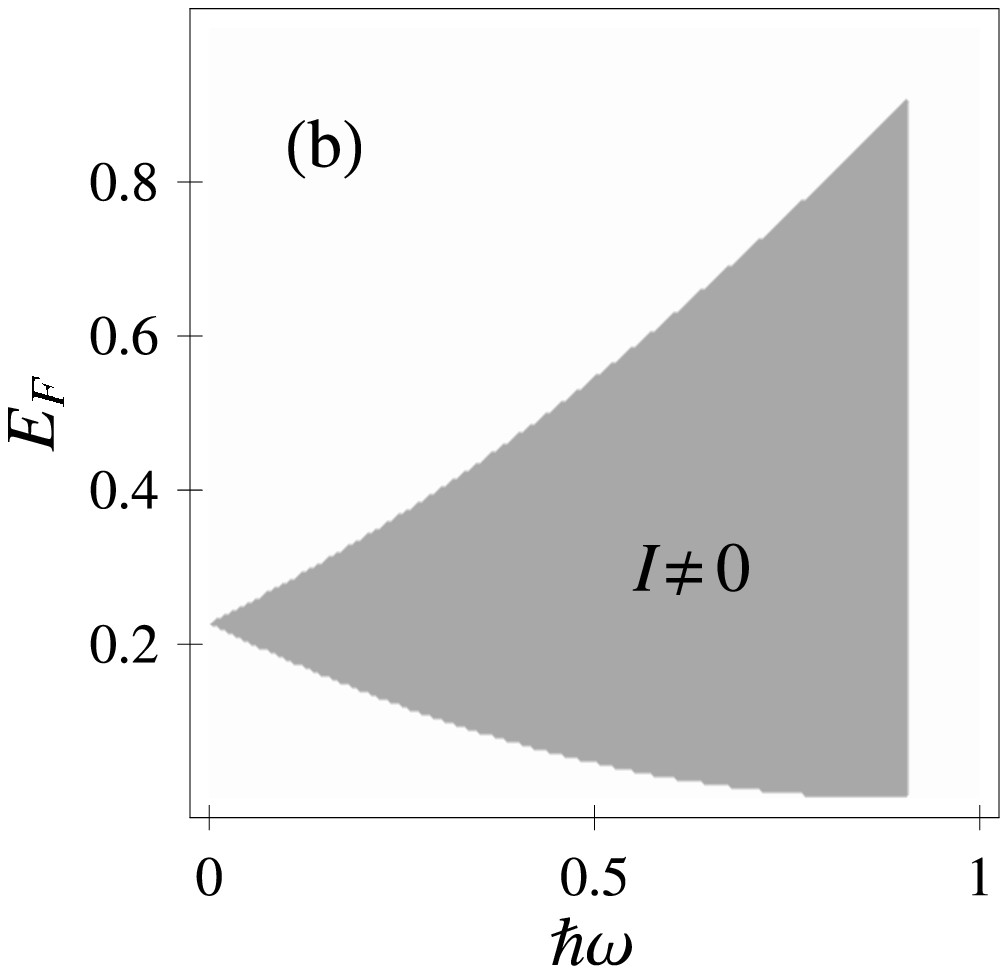}
\caption{(a) Transitions induced by $\sigma_+$ and $\sigma_-$ radiation. (b)
Region of $(E_F,\hbar\omega)$ space with non-zero current. All calculations
presented in this paper have been made using $m^*=0.067m_e$, $R=25$nm and
$E_0=250$V/m. $E_F$ and $\hbar\omega$ are in meV. } \label{figN2}
\end{figure}

The time-dependent potential in Eq.~(\ref{intpot}) induces transitions between
different states. Fig \ref{figN2}(a) shows the transition scheme for an
infinite quantum wire bent into a coil. In this coil geometry there is a full
translational symmetry  with respect to the coordinate $s$, leading to the
exact energy and momentum conservation in the transition. This case need to be
discussed first in order to better understand the physical behavior of the our
structure, in which the translational symmetry is broken due to the presence of
straight and curved segments. For $\sigma_-$ radiation, the exact
energy-momentum conservation conditions read
\begin{eqnarray}
E(k'')=E(k')\pm\hbar\omega, \label{Epm} \\ k''=k'\pm\frac{1}{R}~. \label{kpm}
\end{eqnarray}
Here $k'$ and $k''$ are $k$-vectors of the initial and final electron states,
$\pm$ corresponds to a process of absorption and emission of a photon, and
$1/R$ is the $k$-vector of the sliding potential (\ref{intpot}). In the case of
$\sigma_+$ radiation, $\pm$ in Eq. (\ref{Epm}) should be changed to $\mp$. We
notice that this transition scheme is similar to the transition scheme in
quantum ring illuminated by circularly-polarized radiation \cite{perpierQR}.

\begin{figure}[b]
\centering
\includegraphics[angle=0,width=8.5cm]{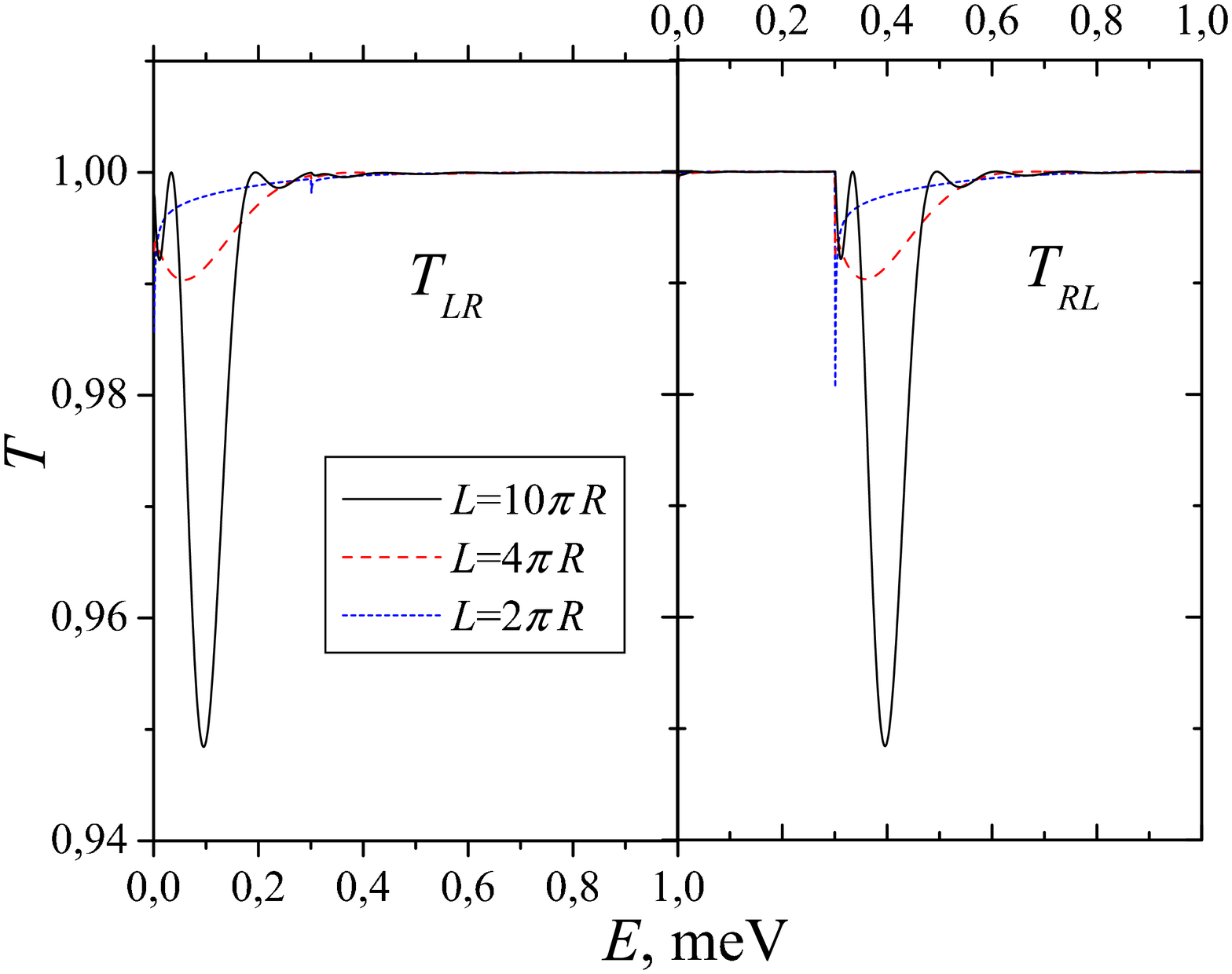}
\caption{Total left-right and right-left transmission probabilities through
curved quantum wires of different curved region lengths $L$ as indicated on the
picture. The circularly-polarized radiation introduces an asymmetry in the
transmission probabilities. In the figure we considered $\hbar\omega=0.3$~meV
and the geometrical potential was neglected.} \label{figN3}
\end{figure}
The key role in the current generation is played by transitions with $k'k''<0$,
i.e. when the direction of electron motion is reversed. Fig. \ref{figN2}(a)
gives examples of such transitions. Without the radiation, the electron
currents due to the left- and right-moving electrons compensate each other.
Radiation-induced transitions with $k'k''<0$ correspond to a change in the
backscattering of only one kind of electrons (for example, the left-moving).
Therefore, the net current becomes non-zero at certain values of the Fermi
energy $E_F$, namely, when the occupation probabilities for the states $k'$ and
$k''$ are different.

From Eqs. (\ref{Epm},\ref{kpm}) we can find that, for a fixed value of the
electromagnetic frequency, the initial and final momentum for the electron
involved in the the transition can be explicitly written as
\begin{eqnarray}
k'=\frac{mR\omega}{\hbar}- \frac{1}{2R}\\ k''=\frac{mR\omega}{\hbar}+
\frac{1}{2R}.
\end{eqnarray}
It is clear that at zero temperature a current can be induced only when
$\hbar^2(k')^2/(2mR^2)<E_F<\hbar^2(k'')^2/(2mR^2)$ and $k'<0$. This can be
satisfied only for some values of the Fermi energy and photon energy. Fig.
\ref{figN2}(b) shows a region of the $(E_F,\hbar\omega)$ space where these
conditions are satisfied.

This analysis applies to the case of a perfect infinite coil-shaped wire. The
energy-momentum conservation for a finite system, such as a curved quantum wire
\cite{perpierBent}, is different than the one defined by Eqs.
(\ref{Epm},\ref{kpm}). In particular, the lack of the full translation symmetry
implies that Eq. (\ref{kpm}) is no longer valid. The numerical calculations
presented in Fig. \ref{figN3} refer to a single curved quantum wire, consisting
of two straight segments smoothly connected by an arc of total length $L$. We
plot there the total transmission  probability for an electron of energy $E$
from the left to the right ($T_{LR}$) and from the right to the left ($T_{RL}$)
in the presence of a $\sigma_{-}$ polarized radiation. The definition of such
probabilities is given below in Eq.~(\ref{te}), and the numerical calculation
is the same discussed in Ref.~\onlinecite{perpierBent}. The figure shows how
the shape of the transmission coefficient is modified by the finite size of the
ring. Instead of a peak at a particular energy determined by the energy
momentum conservation, a broader shape for the transmission coefficient
appears.  Notice that the transmission probabilities $T_{RL}$ and $T_{LR}$ for
the $L=10\pi R$ quantum wire show clear peaks exactly at the
energies defined by Eqs. (\ref{Epm},\ref{kpm}). 
The peak in $T_{LR}$ corresponds to a transition with absorption of a photon
($k'\rightarrow k''$ transition in Fig. \ref{figN2}(a)), the peak in $T_{RL}$
is related to emission of a photon ($k''\rightarrow k'$ transition). By
decreasing $L$ these peaks shift to lower energy and become broader. For
$L=2\pi R$, only small peaks are observed. In this short ring limit, momentum
conservation is released and only the energy conservation condition is
maintained ( i. e. electrons with $E<\hbar\omega$ can not emit $\hbar \omega$
photons). Moreover, the shape of the transmission coefficients in this case
follow the typical $1/\sqrt{E}$ dependence of the one dimensional density of
states.

\section{Calculation method \label{sec2}}
In the ballistic transport regime, the electric current from the left (L) to
the right (R) reservoirs can be written using a generalization
\cite{datta,lit5,galperin} of the Landauer-B\"uttiker formula \cite{landauer}
that takes into account the presence of a radiation field of frequency
$\omega/2 \pi$
\begin{eqnarray}
I=\frac{2e}{h}\sum\limits_n \int\limits_0^\infty \left[
T_{R,L}(E+n\hbar\omega,E)f_{\mu_L}- \right. \nonumber \\ \left.
T_{L,R}(E+n\hbar\omega,E)f_{\mu_R} \right] dE. \label{curr}
\end{eqnarray}
Here $e$ is the electron charge and $T_{R,L}(E+n\hbar\omega,E)$ is the
probability that an electron of energy $E$ in the left reservoir is transmitted
to the right reservoir in a state of energy $E+n\hbar\omega$.  Since we are
going to study the current in the absence of external bias, i.e., at
$\mu_L=\mu_R=\mu$, Eq. (\ref{curr}) can be rewritten as
\begin{equation}
I=\frac{2e}{h}\int\limits_0^\infty  \Delta T(E)f_{\mu}dE,
\label{curr1}
\end{equation}
where
\begin{eqnarray}
\Delta T(E)&=&\sum\limits_n\left[ T_{R,L}(E+n\hbar\omega,E)-
T_{L,R}(E+n\hbar\omega,E) \right] \nonumber\\
&=&T_{RL}(E)-T_{LR}(E). \label{te}
\end{eqnarray}

We will calculate the photocurrent taking into account only single photon
absorption and emission processes. This is equivalent to restrict $n$ to $-1,
0, 1$ values. The wave function in the first region (near the left contact, see
Fig. \ref{fig1}(a)) can then be written as
\begin{equation}
\psi_1=\sum\limits_{n=-1,0,1}\left( A_{n,+}e^{ik_n x}+
A_{n,-}e^{-ik_n x})\right)e^{-\frac{i(E+n\hbar\omega)t}{\hbar}},
\label{wfwire}
\end{equation}
where $k_n=\sqrt{2m^*(E+n\hbar\omega)/\hbar^2}$. The wave function $\psi_2$
near the right contact is given by the same expression with $A$ replaced by
$B$. We use the transfer matrix method~\cite{transfmatr} to calculate the
transmission through the whole system. The total transfer matrix $M$ couples a
vector
$(A)=\left(A_{-1,+},A_{-1,-},A_{0,+},A_{0,-},A_{1,+},A_{1,-}\right)^{\textnormal{T}}$
to a vector $(B)=\left(
B_{-1,+},B_{-1,-},B_{0,+},B_{0,-},B_{1,+},B_{1,-}\right)^{\textnormal{T}}$ as
\begin{equation}
(A)=M\cdot (B). \label{matreq}
\end{equation}
For a system of $N$ rings $M$ can be written as
\begin{equation}
M=\mathcal{M}_0 \left( \prod\limits_{p=1}^{N-1} \mathcal{N}_p
\mathcal{M}_p\right) \mathcal{N}_p, \label{transfmat}
\end{equation}
where $\mathcal{M}_p$ is a transfer matrix of a straight quantum wire segment
$p$ and $\mathcal{N}_p$ is a transfer matrix of the junction of the quantum
wire with the $p$-th ring. We define the segment $0$ as the part of quantum
wire going from the $L$ reservoir to the first ring, the segment $1$ is from
the first ring to the second ring, and so on. The matrix $\mathcal{M}_p$ has a
simple diagonal form
\begin{equation}
\left(\begin{array}{cccccc} e^{-ik_{-1}L_p} & 0 & 0 & 0 & 0& 0
\\ 0 & e^{ik_{-1}L_p} & 0 & 0 &
0 & 0
\\ 0 & 0 & e^{-ik_{0}L_p} & 0 & 0 &
0
\\ 0 & 0 & 0 &  e^{ik_{0}L_p}
& 0 & 0
\\0 &
0 & 0& 0 & e^{-ik_{1}L_p} & 0
\\ 0 &  0 & 0 &
0 & 0 & e^{ik_{1}L_p}
\end{array} \right),
\end{equation}
where $L_p$ is the length of $p$-th segment.

In order to find the transfer matrix $\mathcal{N}_p$ we consider a single
quantum wire-ring junction. The transfer matrix is obtained from the equations
that couple the electron wave function in different branches of the junction.
We formulate these equations using the Shapiro scheme ,\cite{Shapiro,Shapiro1}
which requires that the scattering matrix $S$, connecting the incoming and
outgoing waves at the junction, is unitary. Using the wave function amplitudes
in the wire $(A)$ defined in Eq.~(\ref{wfwire}),and the similarly defined
amplitudes in the ring near the junction in the regions $(B)$, $(F)$ and $(G)$
(see Fig. \ref{fig1}(b)) we couple the in-going and out-going waves in the
junction as
\begin{equation}
\left( \begin{array}{c} A_{i,+}\\B_{i,-}\\F_{i,-}\\G_{i,+}
\end{array} \right)= S  \left( \begin{array}{c}
A_{i,-}\\B_{i,+}\\F_{i,+}\\G_{i,-}
\end{array} \right) \label{smatreq}
\end{equation}
where $i=-1,0,1$ and the $S$-matrix is chosen in the form
\begin{equation}
S= \left( \begin{array}{cccc} 0 & \sqrt{1-a^2} & a & 0
\\ \sqrt{1-a^2} & 0& 0 & a \\ a & 0 & 0 & -\sqrt{1-a^2}  \\ 0 & a
& -\sqrt{1-a^2} &0
\end{array} \right). \label{SMatr}
\end{equation}
Here, the parameter $0<a<1$ describes the splitting of an in-going wave between
different out-going channels. The $a\sim 0$ limit corresponds to a quantum ring
detached from the wire, while the $a\sim 1$ corresponds to a wire with a loop.
The actual value of $a$ depends on the microscopical structure of the junction.
We will consider $a$ as a free parameter in our theory.

The next step is to match at the boundaries the wave function amplitudes $(F)$
and $(G)$ with the solution of the Schr\"odinger equation in the ring. The
procedure has beed described in detail in Ref. \onlinecite{perpierBent}. Here,
we summarize the main steps of the derivation. Let us consider first the
electron wave function in the ring $\psi_r$. The single electron Hamiltonian in
the effective-mass approximation is given by
\begin{equation}
H=-\frac{\hbar^2}{2m^*}\frac{\partial ^2}{\partial
s^2}-\boldsymbol{dE} , \label{ham}
\end{equation}
where $m^*$ is the effective mass, $s$ is a coordinate along the ring. The
dipolar interaction term $\boldsymbol{d}=-e\boldsymbol{r}$ has been introduced
earlier in Eq. (\ref{intpot}). We neglect the geometrical potential
$U_g=\hbar^2/(8m^*R^2)$~\cite{costa81}, which is not expected to affect the
total current in the system. Taking into account only single photon absorption
and emission processes, corresponding to $n=-1,0,1$ in Eq.~(\ref{te}), we can
write the electronic wave function in the form
\begin{equation}
\psi_r(s,t)=\sum\limits_{n=-1}^{1}f_n(s)e^{-\frac{i(E+n\hbar\omega)t}{\hbar}}.
\label{wavef}
\end{equation}
We consider the case of $\sigma_-$ polarization, as shown in Fig. \ref{fig1}.
The $\sigma_+$ case is analogous. By substituting Eq.~(\ref{wavef}) in the
time-dependent Schr\"odinger equation and neglecting multi-photon absorption
and emission terms, we obtain
\begin{eqnarray}
(E-\hbar\omega)f_{-1}+\frac{\hbar^2}{2m^*}f_{-1}''=ieE_0Re^{-i\frac{s}{R}}f_0,
\label{eqf1}\\
Ef_0+\frac{\hbar^2}{2m^*}f_0''=ieE_0R\left(e^{-i\frac{s}{R}}f_1-e^{i\frac{s}{R}}f_{-1}\right),
\label{eqf2}
\\
(E+\hbar\omega)f_1+\frac{\hbar^2}{2m^*}f_1''=-ieE_0Re^{i\frac{s}{R}}f_0.
\label{eqf3}
\end{eqnarray}
By seeking for solutions of Eqs. (\ref{eqf1}-\ref{eqf3}) in the form
$f_{-1}=C_{-1}e^{i\left(\tilde{k}-\frac{1}{R}\right)s}$,
$f_0=C_0e^{i\tilde{k}s}$, and
$f_1=C_1e^{i\left(\tilde{k}+\frac{1}{R}\right)s}$, we obtain the following
system of equations for the coefficients $C_i$
\begin{eqnarray}
(E-\hbar\omega)C_{-1}-\frac{\hbar^2\left(\tilde{k}-\frac{1}{R}\right)^2}{2m^*}C_{-1}-
\nonumber \\ ieE_0RC_0=0, \label{coef1}\\
(E)C_0-\frac{\hbar^2\tilde{k}^2}{2m^*}C_0+ieE_0R\left(C_{-1}-C_1\right)=0,
\label{coef2}
\\
(E+\hbar\omega)C_1-\frac{\hbar^2\left(\tilde{k}+\frac{1}{R}\right)^2}{2m^*}C_1+
\nonumber \\ ieE_0RC_0=0. \label{coef3}
\end{eqnarray}

By imposing that the matrix in the linear system of
Eqs.~(\ref{coef1}-\ref{coef3}) has determinant equal to zero, we obtain six
complex values of $\tilde{k}$. The wave function in the ring can then be
written as
\begin{equation}
\psi_{r}(s,t)=\sum\limits_{j=1}^6\sum\limits_{n=-1}^1C_{n,j}e^{i\left(\tilde{k}_j+n\frac{1}{R}\right)s}
e^{-\frac{i(E+n\hbar\omega)t}{\hbar}}. \label{wavef2}
\end{equation}

Using the continuity of the wave function and the current conservation
condition, we match, in the vicinity of junction, the plane waves defined by
$(G)$ and $(F)$ to $\psi_r$ at $s=0$ and $s=2\pi R$, correspondingly. From the
matching conditions we obtain
\begin{eqnarray}
F_{n,\pm}=\frac{1}{2}\sum\limits_j\left(1\pm
\frac{\tilde{k}_j+\frac{n}{R}}{k_n}\right)C_{n,j}, \\
G_{n,\pm}=\frac{1}{2}\sum\limits_j\left(1\pm
\frac{\tilde{k}_j+\frac{n}{R}}{k_n}\right)e^{i\left(\tilde{k}_j+n\frac{1}{R}\right)2\pi
R}C_{n,j}.
\end{eqnarray}
Moreover, from Eqs. (\ref{coef1}) and (\ref{coef3}) we can express $C_{\pm
1,j}$ as a function of  $C_{0,j}$. Therefore, we obtain $12$ linear equations
given by (\ref{smatreq}) involving 18 variables $A_{i,\pm}$, $B_{i,\pm}$, and
$C_{0,j}$. We can write these equations as
\begin{equation}
\mathcal{K}_p \left( \begin{array}{c}
A_{-1,+}\\A_{-1,-}\\A_{0,+}\\A_{0,-}\\A_{1,+}\\A_{1,-}\\
C_{0,1}\\C_{0,2}\\C_{0,3}\\C_{0,4}\\C_{0,5}\\C_{0,6}
\end{array}
\right)=\left( \begin{array}{c}
\sqrt{1-a^2}B_{-1,+}\\\sqrt{1-a^2}B_{0,+}\\\sqrt{1-a^2}B_{1,+}\\B_{-1,-}\\
B_{0,-}\\B_{1,-}\\0\\0\\0\\aB_{-1,+}\\aB_{0,+}\\aB_{1,+},
\end{array}
\right)\label{longeq}
\end{equation}
where $\mathcal{K}_p$ is a matrix. Multiplying Eq. (\ref{longeq}) by
$\mathcal{K}_p^{-1}$ we can find an expression for  $A_{n,\pm}$ as a function
of $B_{n,\pm}$. The coefficients in these expressions give the transfer matrix
$\mathcal{N}_p$ defined in Eq. (\ref{transfmat}).

The total transfer matrix $\mathcal{M}$ is calculated using Eq.
(\ref{transfmat}). The scattering coefficients for an electron coming, let's
say, from the left contact are found from Eq. (\ref{matreq}) by putting
$(A)=\left(0,A_{-1,-},1,A_{0,-},0,A_{1,-}\right)^{\textnormal{T}}$ and
$(B)=\left( B_{-1,+},0,B_{0,+},0,B_{1,+},0\right)^{\textnormal{T}}$. The total
transmission coefficient from the left to the right reservoir at the energy $E$
is then given by
\begin{eqnarray}
\sum\limits_{n=-1}^1
T_{R,L}(E+n\hbar\omega,E)=|B_{0,+}|^2+\frac{k_1}{k_0}|B_{1,+}|^2+
\nonumber \\ \theta\left(
E-\hbar\omega\right)\frac{k_{-1}}{k_0}|B_{-1,+}|^2,
\end{eqnarray}
where $\theta(z)$ is the step function. The total transmission
from the right to left reservoir is determined in a similar way.

\begin{figure}[tb]
\centering
\includegraphics[width=8.5cm]{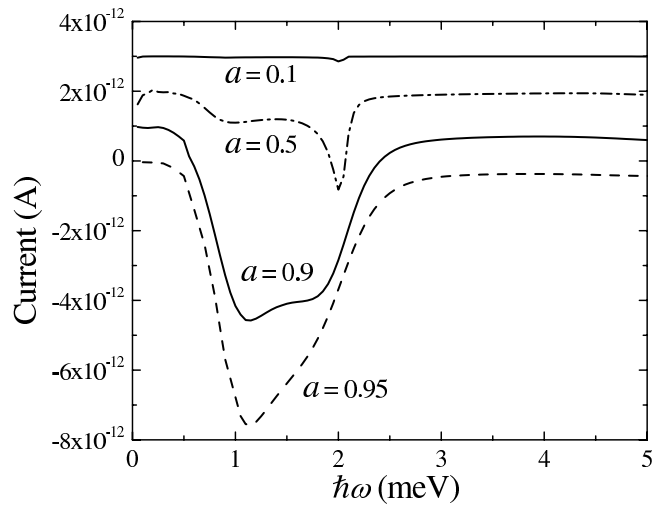}
\caption{Photocurrent in the system with a single ring as a
function of photon energy for different values of the coupling
constant $a$. For all plots, $E_0=250$V/m, $R=25$nm, $\mu=1$meV,
$T=1$K, $m^*=0.067m_e$, and $L=100$nm. All the curves other than
$a=0.95$ have been vertically shifted by steps of $10^{-12}$A for
clarity. } \label{fig2}
\end{figure}

\section{Results \label{sec3}}

Using the above scheme we have calculated the photocurrent as a function of
different parameters. Fig.~\ref{fig2} shows the photocurrent as a function of
photon energy for different values of the coupling parameter $a$ in a structure
with a single ring. In the strong coupling limit ($a\sim 1$) the electron
trajectory involves mainly a single pass in the ring. In fact, according to Eq.
(\ref{SMatr}), an electron is injected with a probability $a^2$ from one part
of quantum wire into the ring and, after traveling within the ring, it procceds
forward into the quantum wire with the same probability. Therefore, in this
limit, the photocurrent dependence on the photon energy is similar to the one
obtained in a bent quantum wire.~\cite{perpierBent} In particular, we observe
that the curve $a=0.95$ is characterized by a dominant peak in the vicinity of
$E_F$.

The presence of this peak at the Fermi energy can be interpreted by considering
the transmission probabilities for left- and right-moving electrons shown in
Fig. \ref{figN3}. The total current is proportional to $T_{LR}(E)-T_{RL}(E)$
integrated up to the Fermi energy. It is clear from Fig.~\ref{figN3} that this
integral would be maximal if $E_F$ is slightly smaller than $\hbar\omega$ since
at this energy only the absorption processes in $T_{LR}$ contribute to the
total current, and the reverse current due to $T_{RL}$ is zero. When the Fermi
energy is bigger than the photon energy there is a contribution from $T_{RL}$
due to the emission of photons that compensate the left to right current. This
leads to a decrease in the current, which eventually reaches zero when all
emission and absorption processes balance $T_{LR}$ and $T_{RL}$.

The photocurrent behavior at smaller values of $a$ is more complex. Notice that
starting at about $a=0.9$ a second peak emerges in the vicinity of $2E_F$. At
$a=0.5$ the peak at $2E_F$ dominates becomes dominant. For smaller $a$ ($a=0.1$
curve) the peak at $E_F$ has practically disappeared. This second peak can be
related to the interference between the two paths of the electron from $(A)$ to
$(B)$: a direct path and a path involving a single or multiple loops in the
ring. Notice also that, as expected, the total photocurrent decreases for
smaller $a$.

\begin{figure}[tb]
\centering
\includegraphics[width=8.5cm]{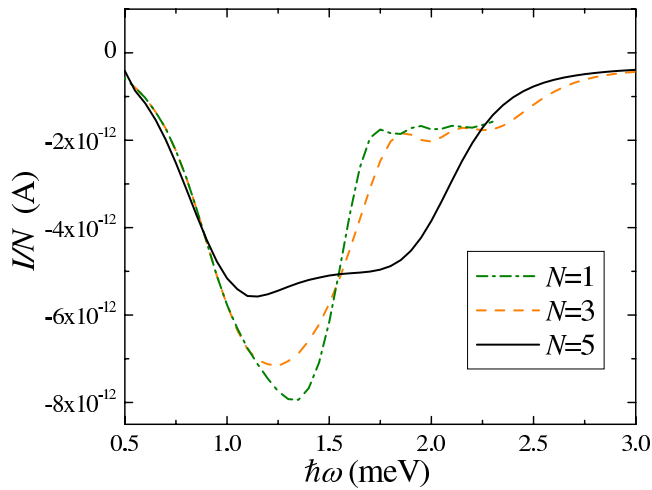}
\caption{Normalized photocurrent $I/N$ as a function of the photon energy for
different number of rings (N) in the system. The coupling coefficient is
$a=0.9$. Notice the strong dependence of the photocurrent profile on the number
of rings.} \label{fig3}
\end{figure}

\begin{figure}[tb]
\centering
\includegraphics[width=8.5cm]{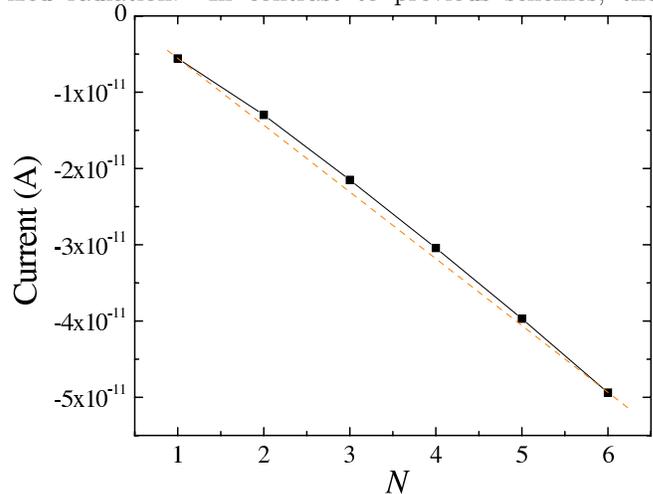}
\caption{Peak value of the photocurrent $I_{peak}$ as a function of the number
of rings $N$ calculated at $a=0.9$. For each value of $N$, the peak value of
photocurrent $I_{peak}$ is determined as a minimum of $I(\hbar\omega)$. The
dashed line connecting the first and last points is a guide for eye. It helps
to notice that the dependence $I_{max}(N)$ slightly deviates from a linear
dependence.} \label{fig4}
\end{figure}

Figs. \ref{fig3} and \ref{fig4} show the main results of the paper. These
figures refer to a strong ring-wire coupling of $a=0.9$, which can give a
strong current. We plot in Fig.~\ref{fig3} the photocurrent normalized to the
number of rings $I/N$ as a function of photon energy for different number of
rings in the system. Notice that photocurrent profile is changing as a function
of $N$, which implies strong nonlinear effects. The normalized amplitude of the
peak in the vicinity of $E_F$ increases and shifts to higher energies with $N$,
while the peak in the vicinity of $2E_F$ decreases. At low photon energies
$\hbar\omega\lesssim 0.7$~meV the photocurrent scales linearly with $N$.

Notice that the peak value of the photocurrent $I_{peak}$ is almost a linear
function of $N$ in the range investigated (the increase of the normalized
amplitude in Fig. \ref{fig3} is related to a positive offset). However, there
are appreciable deviations from a pure linear dependence (see the dashed curve
in Fig. \ref{fig4}). In fact, a fit to the numerical results give a quadratic
dependence, which suggests that multi-ring interference affects the total
current in the system.

\section{Conclusions \label{sec4}}
We have discussed  photocurrent generation in a quantum wire-ring mesoscopic
system by circularly polarized radiation. In contrast to previous schemes, the
present system can be scaled to include a large number of rings, which may lead
to stronger photocurrent. We have shown that the photocurrent increases almost
linearly with the number of rings within a wide range of parameters. Moreover,
we have observed a weak non-linear dependence. From an experimental point of
view, a mesoscopic system with a quantum wire and side-coupled rings can be
fabricated using standard growth methods. This method of current generation
could be explored also in carbon nanotube loops, in which transport properties
have been recently experimentally investigated.~\cite{loops}

\section*{Acknowledgments}

We are grateful to Prof. M. Dykman for many fruitful discussions. This research
was supported by the National Science Foundation, Grant  NSF DMR-0605801.

\end{document}